\documentclass[12pt]{article}
\usepackage{rotating}
\usepackage{epsfig,amssymb}
\usepackage{ulem}
\usepackage{amsmath}
\input{epsf}

\def\laq{\raise 0.4 ex \hbox{$<$}\kern -0.8 em\lower 0.62 ex\hbox{$\sim$}}
\def\gaq{\raise 0.4 ex \hbox{$>$}\kern -0.7 em\lower 0.62 ex\hbox{$\sim$}}

\def\beq{\begin{equation}}
\def\eeq{\end{equation}}
\def\beqa{\begin{eqnarray}}
\def\eeqa{\end{eqnarray}}

 \def\frac#1#2{{\textstyle{{#1}\over {#2}}}}
 \def\lsim{\mathrel{\rlap{\lower4pt\hbox{\hskip1pt$\sim$}}
    \raise1pt\hbox{$<$}}} \def\gsim{\mathrel{\rlap{\lower4pt\hbox{\hskip1pt$\sim$}}
    \raise1pt\hbox{$>$}}}
\def\sqr#1#2{{\vcenter{\vbox{\hrule height.#2pt
         \hbox{\vrule width.#2pt height#1pt \kern#1pt
         \vrule width.#2pt}
         \hrule height.#2pt}}}}

 \def\frac#1#2{{\textstyle{{#1}\over
{#2}}}} 
\def\lsim{\mathrel{\rlap{\lower4pt\hbox{\hskip1pt$\sim$}}
\raise1pt\hbox{$<$}}}
\def\gsim{\mathrel{\rlap{\lower4pt\hbox{\hskip1pt$\sim$}}
\raise1pt\hbox{$>$}}} \def\sqr#1#2{{\vcenter{\vbox{\hrule height.#2pt
\hbox{\vrule width.#2pt height#1pt \kern#1pt \vrule width.#2pt} \hrule
height.#2pt}}}}

\def\beq{\begin{equation}} \def\eeq{\end{equation}}
\def\beqa{\begin{eqnarray}} \def\eeqa{\end{eqnarray}}

\def\gappeq{\mathrel{\rlap {\raise.5ex\hbox{$>$}} {\lower.5ex\hbox{$\sim$}}}}

\def\lappeq{\mathrel{\rlap{\raise.5ex\hbox{$<$}}
{\lower.5ex\hbox{$\sim$}}}}

\usepackage{color}

\begin{document}
\pagestyle{plain}

\begin{flushright}
\end{flushright}
\vspace{15mm}

\begin{center}

{\Large\bf Cooling the Earth with $CO_2$ filled containers in space}

\vspace*{1.0cm}

Orfeu Bertolami$^{1,2}$, and Clovis Jacinto de Matos$^{3}$ \\
\vspace*{0.5cm}
{$^{1}$ Departamento de F\'{\i}sica e Astronomia, Faculdade de Ci\^encias,
Universidade do Porto, \\
Rua do Campo Alegre s/n, 4169-007 Porto, Portugal}\\

{$^{2}$ Centro de F\'{\i}sica das Universidades do  Minho e do Porto,
Rua do Campo Alegre s/n, 4169-007 Porto, Portugal}\\

{$^{3}$ European Space Agency - ESA Headquarters, 75007 Paris, France}

\vspace*{2.0cm}
\end{center}

\begin{abstract}
\noindent
We argue that geostationary (GEO) reflective containers filled with $CO_2$ could be used as shading devices to selectively cool areas on Earth's surface. This proposal would be an interesting addition to the recently discussed suggestion of dumping $CO_2$ to space through the well of a space lift \cite{OB2023}. We also explore the possibility of producing propellants in GEO out of greenhouse gases expelled from the space lift.  Finally, we discuss the much less effective idea of filtering the most prominent infrared bands of the incoming solar radiation using the $CO_2$ wrapped in transparent vessels.
\end{abstract}

\vfill
\noindent\underline{\hskip 140pt}\\[4pt]
\noindent
{E-mail addresses: orfeu.bertolami@fc.up.pt; clovis.de.matos@esa.int}

\newpage

\section{Introduction}
\label{sec:introduction}

In a recent publication \cite{OB2023} it was proposed that $CO_2$ could be transported to space by some suitable adaptations of the space lift. It was shown that the well of a geostationary orbital lift or space elevator as it is usually referred to, could be used for dumping greenhouse gases into space. Naturally, it was assumed that the known requirements to build a stable orbital lift are satisfied, and it was discussed how to use this infrastructure to dump greenhouse gases away from Earth's atmosphere.

Indeed, based on recent advances in material science, more specifically, in the development of carbon nanotubes and macro-scale single crystal graphene, it has been pointed out that the space lift could be a reality  in a foreseeable future (see Ref. \cite{IAA} for a recent assessment). In Ref.  \cite{OB2023}, the use of the space lift well was considered to dump the excess of anthropogenic atmospheric $CO_2$ into space,  but of course, the concept could also be considered for other greenhouse gases (like for example $CH_4$).

The typical dimensions of the orbital lift are its vertical extension, $r_G \simeq 35786$ km, and an assumed constant cross-sectional area, $A=\pi r^2$, where $r$ is the radius of the well. The anchor of the orbital lift could be a geostationary satellite in an equatorial plane orbit. The idea of Ref.  \cite{OB2023} is to inject $CO_2$ into the well of the orbital lift and generate an upward flow so to dump $CO_2$ into space. Natural conditions do not allow for any effective upwards flow as Earth's escape velocity is much greater than the typical average velocities of the $CO_2$ molecules in the air. Therefore, conditions for an upward flow must be engineered. The steps to create an upward flow of $CO_2$ molecules are detailed in Ref.  \cite{OB2023} and, in its simplest form, involve the following steps: i) separation of the $CO_2$ in the air and its injection into the well, which has been emptied of its initial content; ii) ionisation of the $CO_2$ in the well via soft X-ray irradiation; iii) acceleration of the charged $CO_2$ through an electric field along the vertical axis of the orbital lift.

Under reasonable assumptions, it was shown that the outward flow of $CO_2$ is given by, $\Phi = j \pi r^2$, where $j= \rho v_f$, $\rho$ and $v_f$ being the $CO_2$ density and $v_f$ the final velocity at the upper end of the well. For $\rho= 4 \times 10^{-4} ~kg/m^3$ and $r = 15 ~m$ one gets: $\Phi_1 = 4.2~ ton/s$ for the scenario where the first and last sections of the well are under the effect of the electric field and a middle section with no electric field; 
 $\Phi_2 = 3.4~ ton/s$ for the scenario where ionised $CO_2$ is accelerated by an electric field in three sections of the well with two intermediate sections with no electric field \cite{OB2023}. These flows correspond to less than $2 \%$ of the anthropogenic $CO_2$ yield.

Improvements on the estimated yields are possible and were discussed in Ref. \cite{OB2023}. However, the point we would like to make in the present work is that the $CO_2$ flow could be used to fill reflective or transparent containers that would reflect or filter part of the incoming solar radiation at the wavelengths, $4.3~\mu m$ and $15~\mu m,$ making it somewhat ''cooler". For sure, in what concerns filtering, the individual $CO_2$ molecules may re-emit the absorbed radiation after a while, however, the absorbed radiation can be dissipated and re-emission avoided if the concentration of $CO_2$ is sufficiently high, the greenhouse effect. We believe that, even though most of the greenhouse effect is due to the absorption of these infrared wavelengths when solar radiation is reflected back to space by the ground on Earth, the depletion of these wavelengths in the incoming radiation may have an attenuating  effect on the net greenhouse effect.

Before we discuss the details of the present proposal let us point out that the thrust due to the injection of gas into space that is transmitted on the structure of the space lift can be estimated to be about $6.3 \times 10^7~N$ for the first scenario discussed above. However, this undesirable effect can be avoided by adopting the simple solution of directing a symmetric ejection of the $CO_2$ along a direction perpendicular to the axis of the well. The desired  cancellation can be achieved by a radially symmetric set of nozzles perpendicular to the axis of the lift at the top end of the well. In what follows, we shall consider that the delivered $CO_2$ gas at the top of the well is used to fill the reflective or infrared absorbing vessels.

\section{Reflective $CO_2$ filled containers}
\label{sec:reflective}

The main idea here is to use the $CO_2$ ejected by the nozzles at the top of the space lift to fill vessels wrapped with reflective material so to redirect back to space the incoming solar radiation. The shade provided by the vessels will depend on their number and configuration. A sufficiently large set of vessels can cover a considerable area. The vessels can be endowed with additional features, but their main property is their reflective power, which can be quite high.

Even though we consider the vessels to be as simple as possible, they can themselves be endowed with nozzles so to use their $CO_2$ content for propulsion. Once the vessels acquire their final volume, they could be attached to each other, transported away from the space lift and cover a patch over an area where cooling is most needed (like e.g. at the poles). As for the vessels themselves they should be wrapped with a reflective material that ensures gas tightness, is resistant to low/high temperatures and to the interplanetary radiation. The backbone of its structure should be light, resistant and paramagnetic. The properties for these structures can be encountered, for instance, in aluminium alloys widely used in the aerospace engineering. However, as the space lift itself might be built with carbon nanotubes, this material could, in principle, be used for the backbone of the vessels too. The flux of $CO_2$ at the end of the space lift is sufficient to fill a great number of vessels and thus, in the theory, there is no limit to the number of vessels to be attached together and to the shading area they can provide.

\section{Reflective $CO_2$ filled containers in the Inner Lagrange Point (L1)}
\label{sec:toL1}

The shadow provided by the reflective containers in GEO discussed above is essentially of the same order of magnitude of the area covered by the ensemble of containers. In order to shade larger areas it would require either quite large reflecting surfaces in GEO or to move the mirror-like structure away from Earth, beyond GEO orbital radius \cite{Early}. Alternatively, locating a mirror of about 9.4 million km$^2$ at the Lagrange Point (L1) between the Sun and Earth at a distance of about $1.5 \times 10^9~m$ from the Earth centre, one could reduce the solar radiation over the Earth disk by about 1.8$\%$, which would suffice to relief the global warming situation \cite{Roy,Angel}. On the other hand, storing $CO_2$ and also $CH_4$ in GEO, through submitting $CH_4$ to similar manipulations using the space lift as described in Ref. \cite{OB2023} for the $CO_2$, would open new interesting avenues to fuel future space missions, including a mission to drive reflective $CO_2$ containers to L1 as suggested in Ref. \cite{MIT}.

In GEO, solar energy could be used to power the pyrolysis of methane, to produce hydrogen,
\begin{equation}
    CH_4 \, \xrightarrow{\text{\textit{heat}}} \, C \, + \, 2 H_2 \label{chem1}
\end{equation}
and to sustain the Sabatier reaction of $CO_2$ to produce methane and water.
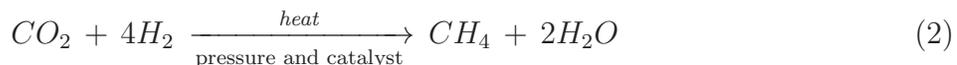
\begin{equation}
    CO_2 \, +\, 4H_2\, \xrightarrow[\text{pressure and catalyst}]{\text{\textit{heat}}} \, CH_4 \, + \, 2 H_2O \label{chem1}
\end{equation}
The electrolysis of water would then allow producing $H_2$ and $O_2$:
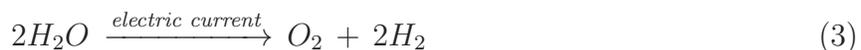
\begin{equation}
    2H_2O\, \xrightarrow{\text{\textit{electric current}}} \,O_2 \, + \, 2H_2 \label{chem1}
\end{equation}
Hence, the $CO_2$ and $CH_4$ greenhouse gases dumped to space by the well of the space lift could allow for yielding and storing important quantities of oxygen, hydrogen and methane propellants directly at GEO. The resulting  infrastructure could fuel all types of space missions and, in particular, a mission to place $CO_2$ filled containers to L1.

In fact, different spacecraft components can, in principle, be brought to assemblage at GEO, so that spacecraft could be assembled and fueled with propellants produced and stored at GEO. Thus, the infrastructure provided by the space lift, and its well, can support the entire logistics of beyond GEO space missions.

The concept proposed here opens the possibility to carry out missions with a challenging high $\Delta v$ cost such as, for instance, Polar orbit ones. This could allow for creating a significant surface to screen Earth's polar caps so to reduce their current melting rate.


\section{Refractive $CO_2$ filled containers}
\label{sec:Refractive}

Let us now discuss the main features of the refractive $CO_2$ filled containers. As the reflective containers,  the refractive containers are filled with $CO_2$ at the top of the space lift. The vessels are essentially transparent and we consider the simplest possible realization of the idea. The most relevant property of the vessels is that they absorb radiation in the relevant infrared wavelengths $4.3~\mu m$ and $15~\mu m$. The absorption effect can be estimated from the radiative equation by the intensity attenuation after crossing a medium with density, $\rho$, and a length, $L$ \cite{Modest}:
\begin{equation}
I_{\nu}(L) = I_0 e^{-\tau_{\nu}},
\label{eq:I}
\end{equation}
where the optical depth of the medium at those specific wavelengths, $\tau_{\nu}$, is given in terms of an integral over a given length:
\begin{equation}
\tau_{\nu} = \int_0^L k_{\nu} \rho ds,
\label{eq:opticald1}
\end{equation}
$k_{\nu}$ being the constant that characterises the medium at a given wavelength.

Assuming that the gas density is uniform, then,
\begin{equation}
\tau_{\nu} = k_{\nu} \rho L.
\label{eq:opticald2}
\end{equation}

The optical depth can also be expressed in terms of microscopic properties through the number density of the absorbing molecules, $n$, and the cross section of the absorbing process, $\sigma$:
\begin{equation}
\tau_{\nu} = n \sigma_{\nu} L.
\label{eq:opticald3}
\end{equation}

It is easy to show that if the $CO_2$ behaves in the vessels as an ideal gas, then:
\begin{equation}
\sigma_{\nu} = k_{\nu} {\mu \over N_A},
\label{eq:sigma}
\end{equation}
where $\mu$ is molecular weight of the $CO_2$ and $N_A$ Avogadro's number.

The necessary data to estimate the $CO_2$ absorption under realistic conditions can be found, for instance, in Ref. \cite{Wei}, in terms of the $CO_2$ concentration. For typical values, that is $400~ppm$, which under PNT conditions corresponds to $\rho_{CO_2} = 5.16 \times 10^{-4}~kgm^{-3}$, one obtains \cite{Wei}: $k \equiv k_{\nu} \rho \simeq 1.0 \times 10^{-1}~m^{-1}$, implying that $k_{\nu} \simeq 1.93 \times 10^{2}~m^2~kg^{-1}$.

The conditions to avoid that the absorbed energy is re-emitted is that the excess energy due to the radiation is thermalised. Assuming a reasonable attenuation figure, say $10\%$ of the incoming radiation on the wavelengths  $4.3~\mu m$ and $15~\mu m$ implies from Eq. (\ref{eq:I}) and (\ref{eq:opticald2}) that $\tau_{\nu} \simeq 10^{-1}$ and with data obtained from Ref. \cite{Wei}, one obtains that $10^{-1}~m^{-1} < k < 6 \times 10^{-1}~m^{-1}$. Still assuming the $CO_2$ in the vessels behaves as an ideal gas, then for the density figure corresponding to $400~ppm$, one obtains the relations between pressure and temperature:
\begin{equation}
\left({p \over 1~atm}\right) = 2.58 \times 10^{-4} \left({T \over 273~K}\right),
\label{eq:pressure}
\end{equation}
As the $CO_2$ freezing point is $T=194.65~K$, we consider, for instance, $T \simeq 200~K$ and hence:
\begin{equation}
\left({p \over 1~atm}\right) = 1.89 \times 10^{-4}.
\label{eq:pressurevalues}
\end{equation}

To evaluate the volume of the vessel, we consider a width, $L=1~m$,  and then the area facing the sun should be $A=10^8~m^2$, meaning that for, say, $100$ vessels, the area facing the sun of each vessel should be $10^6~m^2$, that is, for vessels that have a paving stone shape with a square face, their length should be $1~km$.

Once the set of vessels are positioned, their $CO_2$ content will absorb the incoming infrared radiation and heat up the contained gas, which will increase its pressure. Thus, the vessels should be built with a material that might allow for an increase in volume of an order of magnitude or so. Once the pressure reaches a certain level, a built in set of symmetrically distributed valves would allow for the release of gas into space.

Notice that the above considerations are equivalent to the discussion of the greenhouse effect on a layer of atmosphere where one admits for the incoming solar radiation a strong transmission and weak absorbing coefficient, $\tau_s \simeq 0.9$ \cite{Andrews}.

The infrared absorbing structures that we propose can be an interesting addition to the arsenal of geoengineering space devices to be built in order to face the ongoing climate change crisis and are a logical step forward on the use of the space lift to dump greenhouse gases into space discussed in Ref. \cite{OB2023}. Of course, most of the  greenhouse effect is due to the atmospheric absorption of the solar radiation reflected by Earth's surface. In rough terms, the main contributors to the reflection are soil, snow, vegetation and water. The last three components do not reflect any incoming radiation in the $4.3~\mu m$ and $15~\mu m$ wavelengths \cite{Rast}, meaning that soil reflection is the main component. It is still to be thoroughly studied the effect of depleting the incoming solar radiation from the main wavelengths that cause the greenhouse effect, but it is reasonable to think that it might have a beneficial impact.



\section{Conclusions and Discussion}
\label{sec:Conclusions}

It is consensual  that anthropogenic climate change due to the accumulation of greenhouse gases in the atmosphere is a major civilizational challenge. In order to face the disruptive effects on the patterns of the climate that are already observed, one needs to embark on deep changes on the tenets of the consumption society powered by cheap fossil fuels and built upon the mistaken assumptions about Earth's resources and its capability to admit an indefinite dump
of waste. Naturally, any solution to the problem involves drastic reduction on emissions and radical socio-economic changes.

Given the structural nature and the depth of the changes that must be implemented, geoengineering proposals have been put forward in order to buy us some time. Indeed, severall proposals have been advanced, some of which controverrsial, such as for instance, ocean fertilisation and alkalinity enhancement, albedo enhancement through passive daytime radiative cooling \cite{Zeven1,Wang}, the use of sky-facing thermally-emissive surfaces to radiate heat back into space \cite{Chen,Munday}, stratospheric aerosol injection (SAI), the so-called ``Budyko blancket"  \cite{Budyko,Crutzen,Rash,Lenton}, cloud brightening or a large set of mirrors in the sky to reflect back into space a fraction of the incoming solar radiation (see Ref. \cite{Lenton} for a review).  Other set of ideas involve space reflectors such as a space mirror \cite{Early,Roy} or a myriad of reflecting bubbles \cite{MIT}. Our proposal of reflective vessels filled with $CO_2$ can be seen as a realization of the latter. Of course, geoengineering proposals involve necessarily some degree of negative effects.

Thus, in principle, any device that reduces and reflects back to space the incoming solar radiation might be useful to reduce the amount of infrared radiation trapped in the atmosphere. In the present work we have considered an hypothetical device that involves containers filled with $CO_2$ gas. The vessels can be reflective or transparent so to filter the two main absorbing wavelength of the $CO_2$. The first possibility is much more promising. The device can be coupled with the space lift proposed in a previous study \cite{OB2023}. In the case of reflective vessel, an all spectrum shade is provided. In the case of the refractive vessels, we have shown that an infrared "shadow" can be created by a set of vessels in a geostationary orbit with the net effect of attenuating the incoming solar radiation in the $4.3~\mu m$ and $15~\mu m$ wavelengths if the vessels were filled with $CO_2$. The use of other greenhouse gases, such as for instance methane, could extend the ``blanket" effect for other wavelengths in the infrared.

The possibility to produce and store rocket propellants ($H_2$, $O_2$, $CH_4$) in GEO out of $CO_2$ and $CH_4$ greenhouse gases pumped from the Earth atmosphere could also support space missions to transport the reflective/refractive vessels to polar orbits, or to the Inner L1 Lagrange point. At L1, it becomes possible to shade the entire Earth disk. As a bonus, having a propellant refuelling station in GEO would unlock many possibilities for new innovative space exploitation/exploration missions.

Given the urgency to avoid that the continuous climbing of the concentration of greenhouse gases drive the Earth System (ES) to a Hot House Earth State \cite{Steffen2018}, even the most apparently unrealistic geoengineering proposals are welcome to partially fix the problem. It should be pointed out that the most pessimistic predictions about a possible collapse of the great regulatory ecosystems \cite{Steffen2018} are in agreement with theoretical analyses based on a physical model of the Earth System and the Hot House Earth State as an inevitable outcome given the present intensity of human activities (see e.g. Refs.  \cite{OB-FF1,OB-FF2,OB-FF3,OB-FF4,AB-OB,OB-FF5}).

For sure, the magnitude of the problem asks, first of all, for a deep change on the assumptions of the neoliberal market economy based on the wrong assumption that a finite planet can afford an indefinite period of economical growth without an inevitable collapse. It is becoming increasingly evident that no sustainable future is possible without an encompassing plan for global economic deceleration and a drastic reduction in the use of fossil fuels.







\bibliographystyle{unsrtnat}

\end{document}